\documentclass[twocolumn]{openjournal} 

\usepackage{orcidlink} 
\usepackage[T1]{fontenc}
\usepackage{ae,aecompl}

\usepackage{graphicx}	
\usepackage{amsmath}	
\usepackage{amssymb}	
\usepackage{natbib}

\usepackage{hyperref}
\hypersetup{
	colorlinks=true,
	urlcolor=blue,    
	linkcolor=black,  
	citecolor=blue,   
}

\defcitealias{Pittordis_2018}{PS18}
\defcitealias{Pittordis_2019}{PS19}
\defcitealias{Badry_2021}{ERH}
\defcitealias{Banik_2018_Centauri}{BZ18}

\hypersetup{citecolor=blue, 
            linkcolor=red, 
            menucolor=blue, 
            urlcolor=blue}  

\newcommand{\km}{{~\rm km}}
\newcommand{\s}{{~\rm s}}

\newcommand{\erg}{{~\rm erg}}
\newcommand{\yr}{{~\rm yr}}

\newcommand{\keV}{{~\rm keV}}

\begin{document}

\title{Jet - counter-jet asymmetry in the jittering jets explosion mechanism of supernovae}


\author{Noam Soker\,\orcidlink{0000-0003-0375-8987}} 
\affiliation{Department of Physics, Technion, Haifa, 3200003, Israel;  soker@physics.technion.ac.il}

\begin{abstract}
I identify a point-symmetric morphology composed of three pairs of ears (small lobes) in the X-ray images of the core-collapse supernova remnant (CCSNR) N63A and argue that this morphology supports the jittering jets explosion mechanism (JJEM). The opposite two ears in each of the three pairs of SNR N63A are not equal to each other as one is larger than the other. From the morphology of SNR N63A, I infer that this asymmetry is due to asymmetrical opposite jets at launching.
Namely, the newly born neutron star that launches the jets that explode the star, does it in many cases with one jet more powerful than its counter-jet. I propose that this asymmetry results from that the accretion disk that launches the jets has no time to fully relax during a jet-launching episode. This implies  that the disk has no time to fully relax to a thin disk and it is rather a thick accretion disk. As well, it implies  that if the disk is born with two unequal sides as expected in the JJEM, then during a large fraction, or even all, of the jet-launching episode the two sides remain unequal. I also show that the magnetic reconnection timescale, which is about the timescale for the magnetic field to relax, is not much shorter than the jet-launching episode, therefore the two sides of the accretion disk might have a different magnetic structure.  In most jet-launching episodes,  the unequal sides of the accretion disk launch two opposite  non-relativistic  jets with different energy from each other.  
\end{abstract}


\section{Introduction} 
\label{sec:intro}

In core-collapse supernovae (CCSNe) the inner core collapses to form a neutron (NS), a process that releases a gravitational energy of $\simeq {\rm few} \times 10^{53} \erg$ (e.g., \citealt{Janka2012}). Most of this energy is carried by neutrinos, while the rest explodes the star with explosion energies of $E_{\rm exp} \approx 10^{49} -10^{52} \erg$ (e.g., \citealt{Burrows2013}). Studies in recent years discuss one of two explosion mechanisms that utilize a small fraction of the gravitational energy towards explosion. These are the delayed neutrino explosion mechanism (e.g., \citealt{BetheWilson1985, Hegeretal2003, Janka2012, Nordhausetal2012, Mulleretal2019Jittering, Fujibayashietal2021, Fryeretal2022, Bocciolietal2022, Nakamuraetal2022, Olejaketal2022, Bocciolietal2023, Burrowsetal2023}) and the jittering jets explosion mechanism (JJEM; e.g., \citealt{Soker2010, Soker2020RAA, Soker2022SNR0540, Soker2023gap, PapishSoker2011, GilkisSoker2014, GilkisSoker2016, Quataertetal2019,  ShishkinSoker2021, ShishkinSoker2023, AntoniQuataert2022, AntoniQuataert2023}). 

In the JJEM the explosion is driven by jittering jets. Namely, jets that intermittent accretion disks around the newly born NS launch at typical velocities of $\simeq 10^5 \km \s^{-1}$ (neutrino observations limit the jets in most cases to be non-relativistic, e.g. \citealt{Guettaetal2020}). Accretion of gas with sub-Keplerian specific angular momentum, i.e., below what is required to form a thin accretion belt around the newly born NS, can still launch jets via a magnetized thick accretion disk (e.g., \citealt{SchreierSoker2016}). 
There are $\approx {\rm few}$ to $\approx 30$ jet-launching episodes. In each typical episode, the jets carry a mass of $\approx 10^{-3} M_\odot$. In each episode, the accretion disk launches two opposite jets along an axis that (almost) stochastically varies from one episode to the next. If there is pre-explosion core rotation then the jets' axis variation is not completely stochastic (e.g., \citealt{Soker2023gap}). Also, the back reaction of the jets on the accreted mass acts to have the jets' axis in the same plane that is defined by the jets' axes of the two previous episodes \citep{PapishSoker2014Planar}. The total jet-activity phase lasts for $\simeq 1 - 10 \s$ or even longer in some electron capture CCSNe (\citealt{WangShishkinSoker2024}). Each jet-launching episode lasts for $\tau_{\rm jet} \simeq 0.01-0.1 \sec$, a timescale that is central to the present study. 
 The number of episodes and the value of $\tau_{\rm jet}$ are mainly determined by the fluctuations in the pre-collapse core convective zones and the instabilities above the NS that amplify them (e.g., \citealt{PapishSoker2011, GilkisSoker2015}). 

The angular momentum of the accreted mass in the JJEM starts with stochastic pre-collapse core convection motion (e.g., \citealt{Soker2010, PapishSoker2014Planar, GilkisSoker2015, Soker2019SASI, ShishkinSoker2022, Soker2022Boosting}), or in some cases with envelope convection motion (e.g., \citealt{Quataertetal2019, AntoniQuataert2022, AntoniQuataert2023}), in particular in electron capture CCSNe \citep{WangShishkinSoker2024}. According to the JJEM, the seed angular momentum perturbations of the pre-collapse convective cells are substantially amplified by instabilities between the newly born NS and the stalled shock at $\simeq 100 \km$ (e.g., see discussion by \citealt{ShishkinSoker2021}). 

One prediction of the delayed neutrino explosion mechanism is the existence of failed CCSNe (e.g., \citealt{KurodaShibata2023} for a very recent paper). The JJEM predicts no failed CCSNe (e.g., \citealt{Soker2022Rev} for a review). Recent studies that find no observational evidence for the existence of a large failed CCSN population (e.g., \citealt{ByrneFraser2022, Strotjohannetal2024}) indirectly support the JJEM. 
  
The JJEM implies that in some (but not all) cases several pairs of jets, the last jets to be launched, shape the CCSN remnant (CCSNR) to have a point-symmetric morphology. Namely, a supernova remnant (SNR) structure where there are two or more morphological features each with a counter-morphological feature on the opposite side of the center of the CCSNR. Such point-symmetry is not possible to form by jet-driven explosions that operate only for rare cases of rapidly rotating pre-collapse cores  (e.g., \citealt{Khokhlovetal1999, Maedaetal2012, LopezCamaraetal2013, BrombergTchekhovskoy2016,  Nishimuraetal2017, WangWangDai2019RAA, Grimmettetal2021, Gottliebetal2022, ObergaulingerReichert2023, Urrutiaetal2023a}) because in these cases the jets maintain a fixed axis. Instabilities alone, as in the delayed neutrino explosion mechanism (e.g.,  \citealt{Wongwathanaratetal2015, Wongwathanaratetal2017, BurrowsVartanyan2021, Vartanyanetal2022, Orlando2023}), cannot account for point-symmetric morphologies because instabilities form stochastic morphological features rather than pairs on opposite sides of the center of the explosion. 
 
In three recent studies, I explored the point-symmetric morphology of three CCSNRs and argue that these strongly support the JJEM, SNR 0540-69.3 \citep{Soker2022SNR0540}, the Vela SNR \citep{Soker2023SNRclass}, and possibly SN 1987A \citep{Soker2023NA1987A}. A tentative identification of point symmetry is in the SNR CTB 1 \citep{BearSoker2023RNAAS}. In this paper, I present the point-symmetric morphology of SNR N63A (section \ref{sec:SNRN63A}) as I identify by six ears in its images (e.g., \citealt{Karagozetal2023} for recent images), and use the unequal ears within each pair to infer about the jittering jet properties at launching (section \ref{sec:Jets}).
I summarize this study in section \ref{sec:Summary}. 

\section{Pairs of ears in supernova remnant N63A} 
\label{sec:SNRN63A}

I examine Chandra X-ray observations (e.g., \citealt{Warrenetal2003, Schencketal2016}) and analyze the morphology of the Large Magellanic Cloud SNR N63A, which is a CCSNR (e.g., \citealt{Warrenetal2003, Yamaguchietal2014}). Optical (e.g., \citealt{ChuKennicutt1988}) and radio (e.g., \citealt{Sanoetal2019}) images reveal only small areas of SNR N63A, and do not help the present analysis,  although the radio images, e.g., as presented by  \cite{Dickeletal1993} do hint on the three ears on the northeast. 

\begin{figure}
\begin{center}
\includegraphics[trim=2.6cm 13.1cm 0.0cm 3.0cm ,clip, scale=0.70]{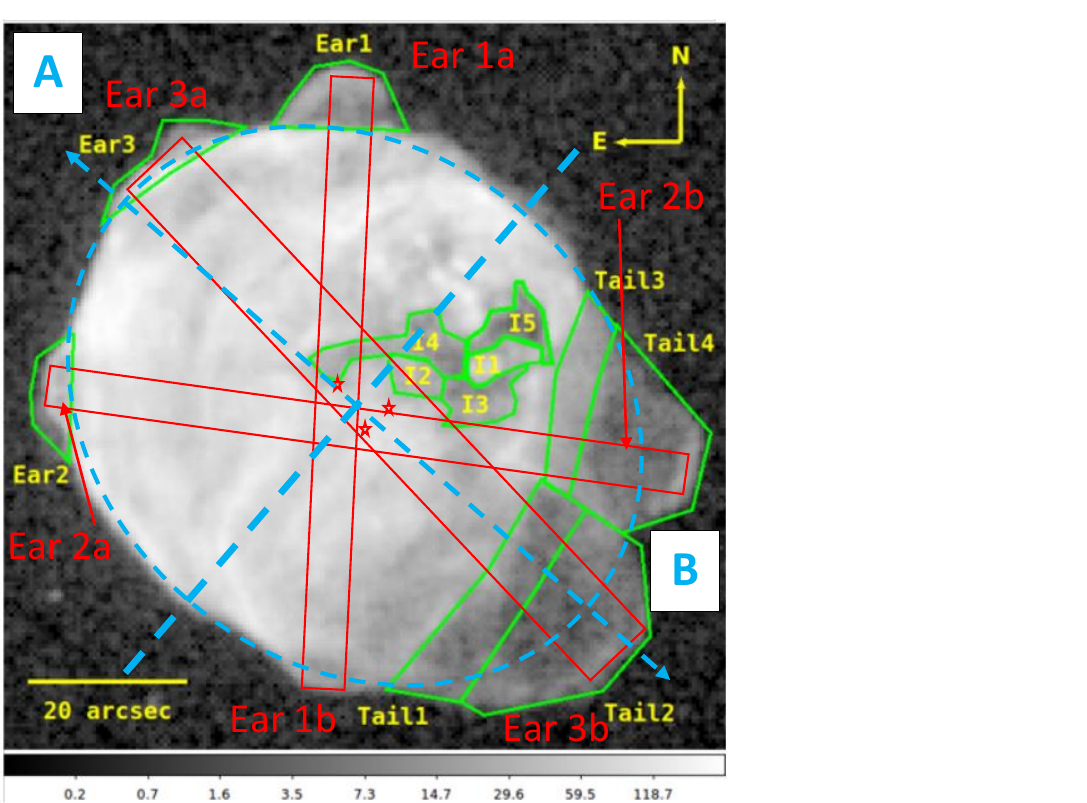}
\end{center}
\caption{A logarithmic-scale $0.3 -7.0 \keV$ X-ray  Chandra image of SNR N63A adapted from figure 2C of \cite{Karagozetal2023}. The green lines and the small-yellow writings are from the original figure of \cite{Karagozetal2023};  the faint regions enclosed by green lines were used for their spectral analysis with the names of the different regions in yellow.  I added the red-solid lines, the dashed-blue lines, and the red and the blue writing.  The red asterisk inside each red rectangle marks its center.   
}
\label{Fig:Figure1} 
\end{figure}

\begin{figure}
\begin{center}
\includegraphics[trim=2.8cm 9.3cm 0.0cm 6.6cm ,clip, scale=0.62]{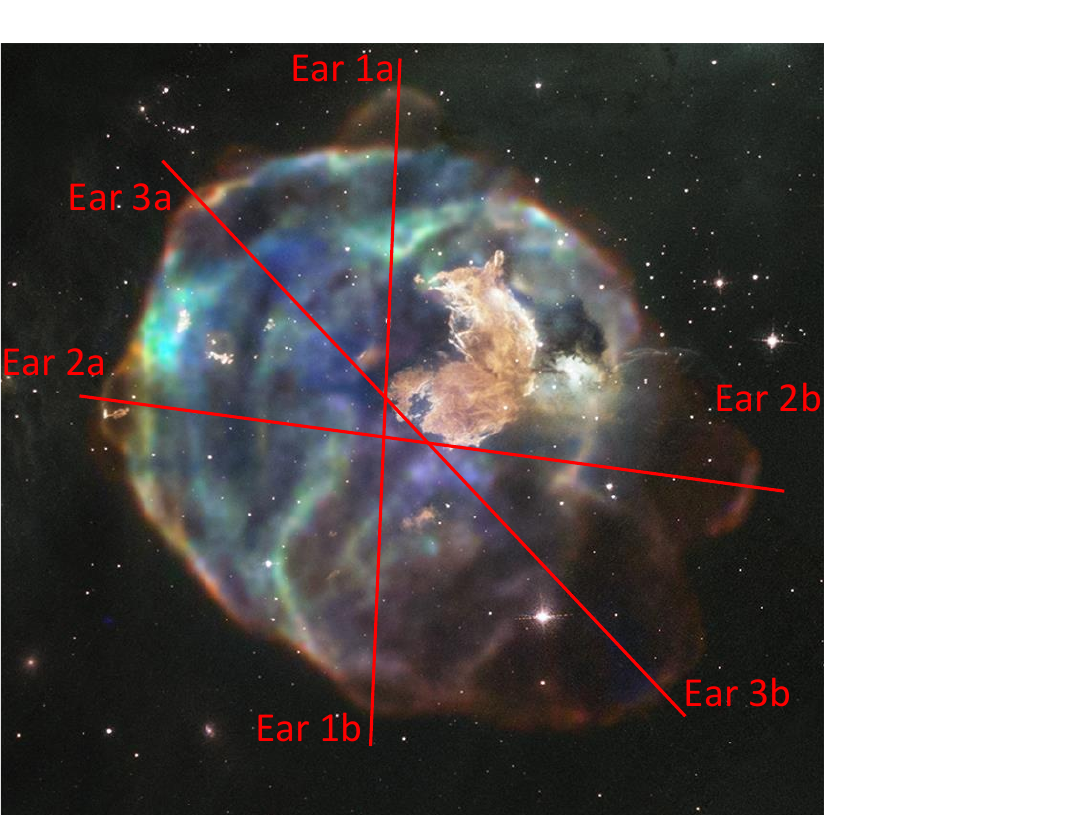}
\end{center}
\caption{A Chandra X-ray (red, green, blue for different X-ray energy bands). The light brown region to the upper right of the three red lines is optical light detected by Hubble.
(Credit: Enhanced Image by Judy Schmidt based on images provided courtesy of NASA/CXC/SAO \& NASA/STScI.)
I added three lines that I drew between the tips of opposite ears. 
}
\label{Fig:Figure2} 
\end{figure}

In Figure \ref{Fig:Figure1} I present an X-ray image from Figure 2C of \cite{Karagozetal2023}. I added the three red rectangles that connect the two ears of the three pairs. The asterisk inside each rectangle marks its center. I label the ears by three pairs 1a-1b, 2a-2b, and 3a-3b. The blue-dashed ellipse is by-eye fitting to the outskirt of the SNR. The thick blue-dashed line along the minor axis of the ellipse splits the plane into region A in the northeast and region B in the southwest. The double-headed blue-dashed line marks what might be the projected (on the plane of the sky) long axis of the SNR. It might more or less coincide with a slow pre-collapse angular momentum of the core because in such a case the jets might jitter around the pre-collapse angular momentum axis. This requires further study. In Figure \ref{Fig:Figure2} I present an X-ray image based on three energy bands, and with an optical image that shows the presence of an ISM or a CSM cloud (e.g., \citealt{Chu2001, Sanoetal2019}). I do not refer to this cloud in the morphological analysis. I added three lines in Figure \ref{Fig:Figure2} that connect the tips of opposite ears.    

Considering that none of the ears has a symmetrical structure around the axis that connects it to its counter ear, it is expected that the centers of the three pairs of ears, as marked by three red asterisks on Figure \ref{Fig:Figure1}, and the center of the ellipse, where the two blue-dashed lines cross each other, do not exactly overlap. Nonetheless, the centers are sufficiently close to each other  (the largest distance between the centers is about $10\%$ of the SNR radius)  to define a point-symmetric morphology for SNR N63A that is based on ears.  

The morphology of SNR N63A reveals the following prominent properties and their relevance to the analysis to follow in section \ref{sec:Jets}. 
\begin{enumerate}
    \item It has a point symmetric structure defined by the six ears (for morphological classification according to shaping by jets see \citealt{Soker2023SNRclass}). This supports the JJEM as instabilities alone and fixed-axis jets do not form such morphologies (section \ref{sec:intro}).  
    \item There is a small signature of elongation along the northeast (region A) to southwest (region B) direction, as indicated by the double-headed blue-dashed line in Figure \ref{Fig:Figure1}.
     Namely, the length of the SNR from Ear 3a to Ear 3b (long diameter of the dashed-blue ellipse) is 1.2 times the diameter of the SNR measured perpendicular to that direction (southeast to northwest which is the short diameter of the ellipse).  As said, this elongation might indicate a slow pre-collapse core rotation around which the late jets that shape the SNR jitter \citep{Soker2023gap}. The same symmetry defines region A where ears 1a-3a reside and region B where ears 1b-3b reside. 
    \item Opposite ears do not have the same size, i.e., the projected area on the plane of the sky $S_{\rm na}$ of ear ${\rm `na'}$ is not equal to the area $S_{\rm nb}$ of its counter-ear, where $n=1,2,3$.  Specifically,  $S_{\rm 1a} \simeq 3 S_{\rm 1b}$, $S_{\rm 2a} \simeq 0.15 S_{\rm 2b}$ and $S_{\rm 3a}  \simeq 0.1 S_{\rm 3b}$.  Two types of processes can cause this asymmetry, either the two opposite jets are unequal at launching or the two sides of the SNR with which equal opposite jets interact are not equal to each other. It does not seem that the effect is due to asymmetrical material in regions A and B because while  
    $S_{\rm 2a} < S_{\rm 2b}$ and $S_{\rm 3a} < S_{\rm 3b}$, for pair number 1 the inequality is $S_{\rm 1a} > S_{\rm 1b}$. I take this to imply that the asymmetry between two opposite ears results from jet-counter-jet asymmetry during the jets' launching process rather than the influence of the medium with which the jets interact. This is the subject of section \ref{sec:Jets}. 
\end{enumerate}

\section{Accounting for unequal pairs of jets} 
\label{sec:Jets}
 This section suggests an explanation for the unequal sizes of ears within each pair of SNR N63A in the frame of the JJEM. 

\subsection{Short jet-launching episodes} 
\label{subsec:ShortEpisode}
The conclusion of section \ref{sec:SNRN63A} is that the three pairs of ears that compose the point-symmetric morphology of SNR N63A were shaped by three pairs of jittering jets wherein each of the jet-launching episodes there was a jet-counter-jet asymmetry. Namely, in the launching process, one jet was more energetic than the other. Many other CCSNRs, in most of them there is only one pair of ears \citep{Soker2023SNRclass}, have unequal opposite ears (e.g, \citealt{Bearetal2017, GrichenerSoker2017ears, Soker2023SNRclass}). 
This jet-counter-jet asymmetry is unlike most other astrophysical objects, from young stellar objects to active galactic nuclei, where in most (but not all) cases the the two opposite jets are about equal to each other, i.e., having no or only a small asymmetry. 
I am discussing here a possible explanation of the unique physical conditions of the JJEM. 

I attribute the asymmetrical jets to the relatively (to other types of systems) small ratio of the total lifetime of the accretion disk to the orbital period of the accretion disk that launches the jets. The accretion disk is extended and therefore I scale the orbital period with a typical radius of the accretion disk around the newly born NS, which has not relaxed yet to its final radius of $\simeq 12 \km$. Each accretion disk that launches one pair of jets in a jet-launching episode in the JJEM lasts for a typical time of $\tau_{\rm jet} \simeq 0.01-0.1 \s$ (section \ref{sec:intro}).  The orbital period of a test particle that orbits the newly born NS, of mass $M_{\rm NS} \simeq 1.4 M_\odot$, at a radius of $r \simeq 20 \km$ is $\tau_{\rm Kep} \simeq 2 \pi r^{3/2}/\sqrt{GM_{\rm NS}} \simeq 0.0013 (r/20 \km)^{3/2} \s$.  I take from now on $M_{\rm NS} \simeq 1.4 M_\odot$ as the analysis is weakly sensitive to the exact NS mass. I neglect relativistic effects which are small, $\simeq 10\%$ effect, compared with the other uncertainties to follow, like the viscosity in the disk and the magnetic reconnection rate. The important ratio in the JJEM is 
\begin{equation}
\left( \frac{ \tau_{\rm jet}}{\tau_{\rm Kep}} \right)_{\rm JJ} \simeq 40 \left( \frac{r}{20 \km} \right)^{-3/2}  \left( \frac{\tau_{\rm jet}}{0.05 \s} \right).   
\label{eq:TauJetKep1}
\end{equation}
For the relevant parameters of the JJEM, this can also be expressed as
\begin{equation}
1 \lesssim \log \left( \frac {\tau_{\rm jet}}{\tau_{\rm Kep}} \right)_{\rm JJ} \lesssim  2 .   
\label{eq:TauJetKep2}
\end{equation}
In most other astrophysical systems the accretion disk lives much longer and the ratio is $\log ( {\tau_{\rm jet}}/{\tau_{\rm Kep}}) \gg 2$.  In young stellar objects even outbursts of high mass accretion rates last for thousands of years (e.g., \citealt{Elbakyanetal2021}), much longer, by five orders of magnitudes, than the Keplerian timescale of about several days at $r \simeq 20 R_\odot$. Radio jets with velocities of $>100 \km \s^{-1}$ from young stellar objects that are launched from near the star last for a year or so (e.g., \citealt{Angladaetal2018}). The Keplerian timescale on the surface of these stars is several hours to a few days. The accretion disk lives longer than the jets. The jets that heat the intracluster medium in clusters of galaxies last of a typical time of $\simeq 10^7 \yr$ (e.g., \citealt{Timmermanetal2022}), compared with the orbital timescale of $\approx 0.01-0.1 \yr$ for an accretion disk around a black hole of $10^{10} M_\odot$.   
The relatively small (with respect to other astrophysical systems) ratio in equation (\ref{eq:TauJetKep2}) explains the jet-counter-jet asymmetry, as I explain now by considering two other timescales that are related to $\tau_{\rm Kep}$. 

\subsection{The viscous relaxation timescale} 
\label{subsec:Viscous}
Consider the supply of material with specific angular momentum $j$ that corresponds to a circular orbit at $r \simeq j^2/GM_{\rm NS}$. The material settled into a more or less steady-state accretion disk on a viscosity timescale. There are large uncertainties in the value of the viscosity in the accretion disk.
 Below I take the relaxation time of a thin-accretion disk. I will find that there is no time for a full relaxation of the intermittent stochastic accretion disks. This implies that the accretion disk will not be thin, but rather a thick disk. In that regard, I note that in some cases the specific angular momentum of the accreted gas might be somewhat lower than what is required for a circular orbit around the newly born NS, i.e., a sub-Keplerian accretion. The JJEM includes the possibility that even sub-Keplerian thick accretion disks can launch jets and that the accretion disk is thick even when its lifetime is longer than the relaxation time \citep{SchreierSoker2016})

In relaxed (stead-state) thin accretion disks where the scale height of the disk at radius $r$ is $H (r)$ and with the usage of the $\alpha$ parameter (e.g., \citealt{ShakuraSunyaev1973}), the viscous timescale is 
\begin{equation}
    \tau_{\rm vis} \simeq \left( \frac{r}{H} \right)^2  \frac{1}{2 \pi \alpha} \tau_{\rm Kep},   
\label{eq:TauViscous1}
\end{equation}
 I scale the ratio $H(r)/r \simeq 0.1$ for a thin accretion disk but I will find that the disk has no time to fully relax to a thin accretion disk.  The viscous time is then 
\begin{equation}
    \tau_{\rm vis} \simeq 0.02 \alpha^{-1} \left( \frac{r}{10 H} \right)^2  \left( \frac{r}{20 \km} \right)^{3/2} \s.    
\label{eq:TauViscous2}
\end{equation}
This leads to the ratio of 
\begin{equation}
0 \la \log \left( \frac{\tau_{\rm jet}}{\tau_{\rm vis}} \right)_{\rm JJ} \lesssim 1. 
\label{eq:RatioViscous}
\end{equation}
This implies that for a relatively long time during a jet-launching episode, or even during the entire episode, the accretion disk is not relaxed.
 This in turn, implies that the accretion disk has no time to relax to a thin disk, and is a thick disk, i.e., $0.3 r \lesssim H(r) \lesssim r$, and the two sides of the disk will not be equal.  If the perturbations, that must exist in the JJEM, form initial two unequal sides of the accretion disk, for a large fraction, possibly for most or even all, of the jet-launching episode they remain unequal. They will therefore launch unequal opposite jets, i.e., one will be more powerful than the other. 

The relation in equation (\ref{eq:RatioViscous}) is in contrast with the majority of other astrophysical objects launching jets for which $\log \left( {\tau_{\rm jet}}/{\tau_{\rm vis}} \right) \gg 1$, and therefore the two sides of the accretion disk have time to relax to the same structure and the same physical conditions with each other during a short fraction of the jet launching episode duration. In those other cases, the accretion disk will launch two symmetrical opposite jets. However, from equation (\ref{eq:RatioViscous}) the jets will not be equal to each other in many cases of the JJEM.     

\subsection{The magnetic reconnection timescale} 
\label{subsec:Reconnection}
Magnetic fields play important roles in the launching process of jets. The rearrangement of magnetic fields involves reconnection. The typical timescale of reconnection is (e.g., \citealt{Parker1979}) $\tau_{\rm rec} \simeq l_{B} / (\epsilon v_{\rm A})$, where $l_{B}$ is the size of the reconnecting magnetic field, $v_{\rm A} = B/\sqrt{4 \pi \rho}$ is the Alfven speed, $\rho$ is the density, and $\epsilon \approx 0.1$. If the accretion disk is supported by the pressure of the magnetic field, namely, $v_{\rm A} > C_s$, where $C_s$ is the sound speed, then the height of the disk is $H(r)\simeq rv_{\rm A}/v_{\rm Kep}$, where $v_{\rm Kep}$ is the orbital velocity of the gas around the newly born NS (I neglect relativistic effects).  The reconnection timescale,  with $v_{\rm Kep} =2 \pi r / \tau_{\rm Kep}$,  is 
\begin{equation}
    \tau_{\rm rec} \simeq \frac{l_B}{\epsilon v_{\rm A}} \simeq 1.6
    \left( \frac{l_B}{H} \right) 
    \left( \frac{\epsilon}{0.1} \right)^{-1} \tau_{\rm Kep}.  
\label{eq:TauRec1}
\end{equation}
If the magnetic pressure is smaller than the thermal plus radiation pressure, the reconnection timescale is even larger.  Note that this derivation does not assume a thin disk. In any case, the expectation is that the disk will not be thin as it does not have time to relax (section \ref{subsec:Viscous}).  

Overall, the reconnection timescale might not be much shorter than the jet-activity period of one jet-launching episode.  With the scaling of equations (\ref{eq:TauRec1}) and  (\ref{eq:TauJetKep1}),  this relation reads 
\begin{equation}
0 \lesssim \log \left( \frac{\tau_{\rm jet}}{\tau_{\rm rec}} \right)_{\rm JJ} \lesssim 1.5. 
\label{eq:RatioRec1}
\end{equation}
This shows that during a large fraction of the jet-launching episode, the accretion disk might not relax to have a symmetrical magnetic structure on both its sides.  This is different from most other astrophysical systems where $\log (\tau_{\rm jet}/ \tau_{\rm rec}) \gg 2 $ and for which the two sides of the accretion disk reach symmetrical magnetic fields with respect to each other  (this inequality is obtained by taking $\tau_{\rm rec} \simeq \tau_{\rm Kep}$ for the young stellar objects and the active galactic nuclei that are mentioned in section \ref{subsec:ShortEpisode}).    

\section{Summary} 
\label{sec:Summary}
 
I analyzed the X-ray morphology of the CCSNR N63A and concluded that it possesses a point-symmetric structure of three pairs of ears. I also argued in section \ref{sec:SNRN63A} that the only viable explanation is the JJEM. Namely, N63A exploded by jittering jets. 
In that, CCSNR N63A is the fourth CCSNR with point-symmetric morphology; the others are SNR 0540-69.3 \citep{Soker2022SNR0540}, the Vela SNR \citep{Soker2023SNRclass}, and possibly SN 1987A \citep{Soker2023NA1987A}. All these strongly support the JJEM and pose challenges to the delayed neutrino explosion mechanism. 

The opposite two ears in each of the three pairs of SNR N63A are not equal to each other. This is a common property of opposite pairs of ears in CCSNRs, including those with only one pair of ears (e.g., \citealt{Soker2023SNRclass}). The ears in a pair are not equal to each other in one or more of their properties, including size, intensity, and distance from the center (e.g., \citealt{Soker2023SNRclass}). I used the morphology of SNR N63A to infer that this asymmetry is due to asymmetrical opposite jets at launching, rather than the unequal environment with which the jets interact as they inflate the ears.  
 The unequal power of the two opposite jets that I inferred from the study of SNR N63A in section \ref{sec:SNRN63A} motivated the study in section \ref{sec:Jets}, where  
I proposed an explanation for the observation that in most CCSNRs the two ears (or lobes) in a pair are not equal to each other. 

I proposed that the different powers of the two opposite jets at launching results from that the accretion disk that launches the jets having no time to fully relax during a jet-launching episode.  Firstly, it implies that the accretion disk stays thick and has no time to fully relax to a thin accretion disk. Secondly,  this implies that if the disk is born with two unequal sides, then during a large fraction, or even all, of the jet-launching episode the two sides remain unequal. In the JJEM that is based on instabilities and angular momentum fluctuations of the accrete gas, each accretion disk is likely born with unequal sides. I showed that the viscous relaxation timescale is not much shorter than the jet-launching episode (equation \ref{eq:RatioViscous}). I also showed that the magnetic reconnection timescale, which is about the timescale for the magnetic field to relax, is not much shorter than the jet-launching episode in the JJEM (equation \ref{eq:RatioRec1}). 

I propose that the two unequal sides of the accretion disk in each jet-launching episode are likely to launch unequal jets that in turn inflate unequal ears. This assertion requires further study. 

Overall, the morphology of CCSNR N63A supports the JJEM and motivates an explanation for the common occurrence of unequal opposite ears in CCSNRs. 

\section*{Acknowledgements}
 I thank an anonymous referee for useful comments.  This research was supported by a grant from the Pazy Research Foundation.


\end{document}